\documentclass[twocolumn,secnumarabic,amssymb, nobibnotes, aps, prd, showpacs,amsmath]{revtex4}
\usepackage[dvipdf]{graphicx}

\usepackage{graphicx}% Include figure files
\usepackage{dcolumn}% Align table columns on decimal point
\usepackage{bm}% bold math
\usepackage{color}
\begin{document}
\title{Amplitude death in complex networks  induced by environment}
\author{V. Resmi}
\email{v.resmi@gmail.com}
\affiliation{Indian Institute of Science Education and Research, 
Pune - 411021, India} 
\author{G. Ambika}
\email{g.ambika@iiserpune.ac.in}
\affiliation{Indian Institute of Science Education and Research, 
Pune - 411021, India} 
\author{R. E. Amritkar}
\email{amritkar@prl.res.in}
\affiliation{Physical Research Laboratory, Ahmedabad - 380009, India}
\author{G. Rangarajan}
\email{rangaraj@math.iisc.ernet.in}
\affiliation{Department of Mathematics and Centre for Neuroscience, Indian Institute of Science, Bangalore 560012, India}
\begin{abstract}
We present a mechanism for amplitude death in coupled nonlinear dynamical systems on a complex network having interactions with a common environment-like external system. We develop a general stability analysis that is valid for any network topology and obtain the threshold values of coupling constants for the onset of amplitude death. An important outcome of our study is a universal relation between the critical coupling strength and the largest non-zero eigenvalue of the coupling matrix. Our results are fully supported by the detailed numerical analysis for different network topologies. 
\end{abstract}
\maketitle
\section{INTRODUCTION}
The dynamics of most real world systems is very complex and can be analyzed by considering them as many sub-systems or units interacting with each other. Such systems then can be modeled by complex networks of diverse topologies chosen to suit their collective behavior. It is now well established that the interaction among the sub-systems can lead to emergent phenomena like synchronization and amplitude or oscillator death \cite{Pikovsky2003,Kaneko1993,Ott1993}. In such cases, the global dynamics depends on the interplay between the network structure and nodal dynamics. However, the interactions of complex networks with environment, such as an external agency or medium, and the consequent emergent dynamics are not studied yet. In the case of quantum systems, the role of environment in causing decoherence, relaxation, and dissipation is well studied \cite{Amm98,Zur93,Gur03,Bra03}. In the context of biological systems, the environment can play a  constructive role as the mechanism for triggering or signaling coordinated rhythms \cite{Kuz05, Wang05,Gon05}. Hence it is important and relevant to study the effect of environment on the dynamics of coupled systems.
  
Among the emergent phenomena in coupled systems, the suppression of dynamics or amplitude death is often a useful control mechanism for stabilizing systems to steady states. It is an important self organized behavior that can play crucial roles in regulating, switching, and controlling  physical \cite{Her00,Wei07}, chemical \cite{Dol88,Cro89,Bar85,Dol96}, and biological systems \cite{Ata06, Kos10, Ozd04, Ull07}. In this context, amplitude death (AD) refers to  the phenomenon where the  coupled or interacting sub-systems settle to a steady state in which dynamics is quenched.

In the context of regularly coupled systems, it has been shown that amplitude death can be induced by different mechanisms, such as parameter mismatch \cite{Bar85, Erm90,Zhai04}, time-delay coupling \cite{Ram98,Pra05,Choe07,Dod04,Atay03}, conjugate coupling \cite{Kar07,Das10}, attractive and repulsive couplings \cite{Chen09}, and dynamical coupling \cite{Kon03}. In addition to the study of amplitude death in regularly coupled systems, there has been a few recent studies on amplitude death in complex networks. In the specific context of networks, amplitude death has been studied for parameter mismatch or detuning of frequencies in an ensemble of limit cycle oscillators with mean field coupling \cite{Mat90}, array of limit cycle oscillators \cite{Rub00,Rub02,Yang07},small-world networks \cite{Hou03}, and scale-free networks \cite{Liu09}. So also, time-delay in coupling is found to induce amplitude death in networks of limit-cycle oscillators \cite{Dod04,Sen05,Kon04,Song11} and chaotic systems \cite{Kon07, Zou11}. Recently, it was shown that it is possible to target amplitude death in a network of nonlinear oscillators by a proper choice of nonlinear coupling \cite{Pra10}.

In this paper, we present an interesting phenomenon in which the collective dynamics of coupled systems is quenched due to an interaction with an environment or external agency.  For this, we model the average effect of the environment by an over-damped oscillator which is kept alive with feedback from the subunits. We find that while the coupling among the units can give rise to a synchronizing tendency, the coupling through the environment has a tendency to drive the systems to a state where the sum of the variables is small. The combined effect of these two tendencies is to lead the coupled systems to the state of amplitude death. Our method has the advantage that it involves a single damped dynamical system coupled to all nodes equally and hence the design procedure is simple and easy to implement. It is found to be effective in complex networks of different topologies. 

We have found this mechanism to be quite general and effective in inducing amplitude death in two systems coupled by different types of coupling and of different intrinsic dynamics \cite{Res11}.  The present study generalizes the previous results in two directions. First, we consider a complex network of $N$ systems and develop the stability analysis following the approach given in Ref.~\cite{Ran02}. The stability conditions are obtained for the general case. Second, we consider different network structures.  Our results are supplemented by detailed numerical analysis where indices of amplitude death are computed directly from simulations. For numerical simulations, we use R\"ossler as a standard system in its chaotic region. However, we have tried this method for Landau-Stuart oscillators and the Hindmarsh-Rose model of neurons and found it to be effective in causing amplitude death. An important finding from our study is that, the critical strength of coupling needed for amplitude death has a universal relation with the largest non-zero eigenvalue of the coupling matrix which is tested for many symmetric networks like chain, ring, tree, lattice, all-to-all, star, and random topologies. 
\section{Amplitude death via direct and indirect coupling}
\label{sec:mech}
We consider the dynamics of $N$ systems $x_i, i=1,2,\ldots,N$, in a network, coupled with two types of couplings, namely, a direct diffusive coupling and an indirect coupling through an environment as an extension of the model given in Ref.~\cite{Res11} which gave a general model for amplitude death in two coupled systems. The dynamics of such a model is given by
\begin{eqnarray}
\dot{x}_i &=& f(x_i) + \sum_{j} \beta G_{ij} \epsilon_d x_j + \epsilon_e \gamma w \nonumber \\
\dot{w} &=& -\kappa w - \frac{\epsilon_e}{N} \gamma^{T} \sum_{i}{ x_{i}}
\label{eq:model}
\end{eqnarray}
where $i,j = 1,2,...N$. Here, $x_i$ represents $m$-dimensional nonlinear oscillators whose  intrinsic dynamics is given by $f(x_i)$. $G$ is the coupling matrix of dimension $N \times N$. We choose the elements of $G$ such that, the row-sum, $\sum_{j} G_{ij} = 0$, for every $j$. This ensures that the largest eigenvalue of the coupling matrix $\mu_1$, is zero. $\beta$ is a matrix ($m\times m$) with elements $0$ and $1$ and defines the components of $x_i$ which take part in the coupling.  For simplicity, we take $\beta$ to be diagonal, $\beta=diag(\beta_1,\beta_2,\ldots,\beta_m)$ and in numerical simulations, only one component, $\beta_1$ is assumed to be non-zero. The environment is considered to be a one-dimensional over-damped oscillator $w$, with damping parameter $\kappa$. It is clear that without feedback from the systems, the environment can not remain dynamic and will rapidly settle to a steady state. However, the feedback from all the systems keeps it active. All the systems, in turn, get feedback from the environment.  $\gamma$ is a column matrix ($m \times 1$), with elements $0$ or $1$, and it decides the components of $x_i$ that get feedback from the environment. $\gamma^{T}$ is the transpose of $\gamma$ and decides the components of $x_i$ which gives feedback to the environment. The strength of this feedback coupling between the systems and the environment is given by $\epsilon_e$. 

\begin{figure}
\includegraphics[width=0.95\columnwidth]{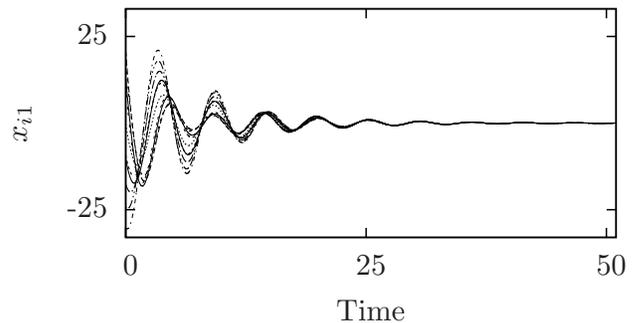}
\caption{ \label{ts:rossler} Time series of the first variables $x_{i1}$ of $10$ coupled R\"ossler systems in an all-to-all coupled network [Eq.~(\ref{eq:rossler})] showing amplitude death for $(\epsilon_d,\epsilon_e) = (0.05,0.8)$. Here, the R\"ossler parameters are $a = b = 0.1, c=18$. The damping parameter of the environment is taken to be $\kappa=1$. }
\end{figure}

We illustrate our scheme using a network of coupled chaotic R\"ossler systems  represented by the following equations:
\begin{eqnarray}
\dot{x}_{i1} & = & -x_{i2} - x_{i3} + \epsilon_d \sum_{j} G_{ij} x_{j1} + \epsilon_e w ,
\nonumber \\
\dot{x}_{i2} & = & x_{i1} + a x_{i2} ,
\nonumber \\
\dot{x}_{i3} & = & b + x_{i3} ( x_{i1} -c ) ,
\nonumber \\
\dot{w} & = & -\kappa w - \frac{\epsilon_e}{N} \sum_{i}{ x_{i1}} .
\label{eq:rossler}
\end{eqnarray}
Here, we choose $G$ to be an all-to-all connected network of $10$ nodes, that is, $G_{ij}=1$, if $j \neq i$ and  $G_{ii}=-9$. We find that amplitude death is possible for suitable values of coupling strengths. The time series for the amplitude death state is shown in Fig.~\ref{ts:rossler}.
We note that this state corresponds to the stable fixed point of  Eq.~(\ref{eq:rossler}), given by
\begin{eqnarray}
\label{eq:rosfp}
x_{i1}^{*} & = & [ c - \sqrt{c^2-  4ab\kappa / (\kappa-\epsilon_e^{2}a) }]/2 , \nonumber \\
x_{i2}^{*} & = & -x_{i1}^{*}/a ,\nonumber \\
x_{i3}^{*} & = & -b/(x_{i1}^{*}-c), \nonumber \\
w^{*} & = & -\epsilon_e x_{i1}^{*}/ \kappa . 
\end{eqnarray}
As noted from Eq.~(\ref{eq:rossler}), our model has two types of coupling. The first is direct diffusive coupling which tends to synchronize the systems $x_{1} = x_{2} = x_3 = \ldots = x_N$. We find that, the coupling via environment has a tendency to decrease the sum, $\sum_i x_i$. When both these tendencies work together the systems converge to a fixed point. This is explicitly seen in the context of two systems coupled through the environment \cite{Res11}, where the environmental coupling reduces the sum to a small value, corresponding to antiphase synchronization.  
\section{Linear stability analysis}
The stability of the steady state of the network of coupled systems given in Eq.~(\ref{eq:model}) can be analysed by writing the variational equations formed by linearising Eq.~(\ref{eq:model}) as 
\begin{eqnarray}
\dot{\xi}_i & = & f'(x_i) \xi_i + \sum_j \beta G_{ij} \epsilon_d \xi_j  + \epsilon_e \gamma z ,
\nonumber \\
\dot{z} & = & -\kappa z - \frac{\epsilon_e}{N} \gamma^{T} \sum_{i} \xi_i ,
\label{eq:deviation}
\end{eqnarray}
where $\xi_{i}$ and $z$ are small deviations from the respective values of $x_i$and $y$, and $f'$ is the $m \times m$ Jacobian matrix.

Let us introduce the $m \times N$ state function \cite{Ran02}
\begin{equation} 
\Xi=(\xi_1,\xi_2,\ldots,\xi_N).
\end{equation}
Then, Eq.~(\ref{eq:deviation}) for the synchronized state ($x_1 = x_2 = \ldots = x_N$) can be written as 
\begin{eqnarray}
\label{eq:row-form}
\dot{\Xi} = f' \Xi + \beta \epsilon_d \Xi G^{T} + \epsilon_e z \Gamma , \\
\label{eq:row-form_z}
\dot{z} = -\kappa z - \frac{\epsilon_e}{N} \gamma^{T} \sum_i \xi_i ,
\end{eqnarray}
where $G^T$ is the transpose of the coupling matrix, and $\Gamma$ is an $m \times N$ matrix, $\Gamma = (\gamma,\gamma,\ldots,\gamma)$ .

Let $e_k$ be an eigenvector of $G^{T}$ such that 
\begin{equation}
G^{T} e_k = \mu_k e_k ,
\end{equation}
where $\mu_k$ is an eigenvalue of $G^T$. 
Right-multiplying both sides of Eq.~(\ref{eq:row-form}) with $e_k$, we get
\begin{equation}
\label{eq:Xiek}
\dot{\Xi} e_k = f' \Xi e_k + \mu_k \beta \epsilon_d \Xi e_k + \epsilon_e z \Gamma e_k .
\end{equation}
Let 
\begin{equation}
\Phi_k = \Xi e_k .
\end{equation}
Then, Eq.~(\ref{eq:Xiek}) can be written as
\begin{equation}
\label{eq:Phik}
\dot{\Phi}_k = f' \Phi_k + \mu_k \beta \epsilon_d \Phi_k + \epsilon_e z \Gamma e_k .
\end{equation}
We note that $e_1 = (1,1,\ldots,1)^{T}$ is the synchronization manifold and $\Phi_1 = \Xi e_1 = \sum_i \xi_i$. Since one could write $\Gamma$ as the product $\Gamma = \gamma e_1^T$, Eqs.~(\ref{eq:Phik}) and (\ref{eq:row-form_z}) can be written as
\begin{eqnarray}
\label{eq:Phik_proper}
\dot{\Phi}_k & = & f' \Phi_k + \mu_k \beta \epsilon_d \Phi_k + \epsilon_e z \gamma e_1^T e_k ,\\
\label{eq:zdot}
\dot{z} & = & -\kappa z - \frac{\epsilon_e}{N} \gamma^{T}  \Phi_1 .
\end{eqnarray}

First, we consider the case where $G$ is taken to be a symmetric matrix. In this case, the remaining eigenvectors span an $(N-1)$-dimensional subspace orthogonal to the eigenvector $e_1$ . Consequently, this subspace is orthogonal to the synchronization manifold. For $k = 1$, Eq.~(\ref{eq:Phik_proper}) becomes
\begin{equation}
\label{eq:Phi1dot}
\dot{\Phi}_1 = f' \Phi_1 + \epsilon_e z N \gamma .
\end{equation}
Since $e_i$ are orthogonal, $\Gamma e_k = \gamma e_1^T e_k = 0$ for $k \neq 1$. Therefore, For $k \neq 1$, Eq.~(\ref{eq:Phik_proper}) reduces to
\begin{equation}
\label{eq:Phikdot}
\dot{\Phi}_k = f' \Phi_k + \mu_k \beta \epsilon_d \Phi_k .
\end{equation}

We note that Eqs.~(\ref{eq:Phi1dot}) and (\ref{eq:zdot}) are coupled while Eq.~(\ref{eq:Phikdot}) is independent of the other two. Moreover, Eq.~(\ref{eq:Phikdot}) is equivalent to the master stability equation introduced by Pecora and Carrol in Ref.\cite{Pec98}. Therefore, the stability function for any given system will be obtained as a function of $\epsilon_d \mu_k$ in the same way. This therefore ensures the stability of the synchronized state $x_1=x_2=x_3=\ldots=x_N$. As noted in the previous section, for the amplitude death state to be stable, we need one more condition to be satisfied. That is, the synchronized state should be a fixed point. For this, the eigenvalues of the Jacobian corresponding to the coupled system given in  Eqs.~(\ref{eq:Phi1dot}) and (\ref{eq:zdot}) should be negative.

So far, we have discussed the case where the coupling matrix $G$ is symmetric. The same analysis can be extended to the asymmetric case as well. In this case, the eigenvectors of $G$ are, in general, not orthogonal to the synchronization manifold.

Let $e_k$, for $k \neq 1$, be split to two components, one parallel and the other perpendicular to $e_1$. That is,
\begin{equation}
\label{perp_par}
e_k = e_k^{\perp} + q_{\parallel} e_1 ,
\end{equation}
where, $e_{k}^{\perp}$ is orthogonal to $e_1$. Substituting $e_k$ from Eq.~(\ref{perp_par}) in Eq.~(\ref{eq:Phik}), we find that the dynamics of $\Phi_1$ and $z$ will again be the same as given in Eqs.~(\ref{eq:Phi1dot}) and (\ref{eq:zdot}) . The dynamics of $\Phi_k, k > 1$, are given by
\begin{equation}
\dot{\Phi}_k = f' \Phi_k + \mu_k \beta \epsilon_d \Phi_k + \epsilon_e z N \gamma q_{\parallel} , 
\label{asym_Phikdotz}
\end{equation}
since $\Gamma e_k^{\perp} = \gamma e_1^T e_k^{\perp} = 0$.
In principle, the coupled equations, Eqs.~(\ref{eq:Phi1dot}), (\ref{asym_Phikdotz}) and (\ref{eq:zdot}) can be considered as a master stability equation in this case also. However, in this case it is not practically useful since the master stability function will be a function of four parameters $\epsilon_d$, $a$, $b$, and $\epsilon_e$, with $\mu_k = a + i b$.
 
To continue the analysis of the stability of the amplitude death states from Eqs.~(\ref{eq:Phi1dot}), (\ref{eq:zdot}), and (\ref{eq:Phikdot}), we assume that the time average values of $f'$ are approximately the same and can be replaced by an effective constant value $\alpha$. In this approximation we treat $\xi_i$'s to be scalars. This approximation simplifies the problem such that only the relevant features remain and is expected to give features near the transition. This type of approximation was used in Refs.~\cite{amb09,Res10,Res11} and it is found to describe overall features of the phase diagram reasonably well.

Thus, Eq.~(\ref{eq:Phikdot}) becomes
\begin{equation}
\label{eq:Phikdot_simpli}
\dot{\Phi}_k = \alpha \Phi_k + \mu_k \beta \epsilon_d \Phi_k ,
\end{equation}
and the corresponding Lyapunov exponent is given by
\begin{equation}
\label{eq:lyap1}
\lambda_1 = \alpha + \mu_2 \epsilon_d ,
\end{equation}
where $\mu_2$ is the largest $\mu_k$ for $k \neq 1$.

The Jacobian corresponding to the coupled Eqs.~(\ref{eq:Phi1dot}) and ~(\ref{eq:zdot}) is 
\[ J =  \left( \begin{array}{cc}
\alpha & \epsilon_e N \\
-\epsilon_e/N & -\kappa \end{array} \right) ,\] 
and the eigenvalues are
\begin{equation}
\lambda_{2,3} = \frac{(\alpha - \kappa) \pm \sqrt{ (\kappa - \alpha)^2 - 4 \epsilon_e^2 } }{2} .
\label{eq:lyap23}
\end{equation}

For the stability of the amplitude death state,  the real parts of the eigenvalues should be negative. Thus Eq.~(\ref{eq:lyap1}) gives the condition 
\begin{equation}
\alpha +  \mu_2 \epsilon_d < 0 ,
\label{eq:stability1}
\end{equation}
while from Eq.~(\ref{eq:lyap23}) we get the following conditions. \\
(1) If $(\kappa - \alpha)^2 < 4 (\epsilon_e^2 - \alpha \kappa )$, $\lambda_{2,3}$ are complex and the condition of stability is 
\begin{equation}
\kappa > \alpha .
\label{eq:stability2_simpli}
\end{equation}
(2) If $(\kappa-\mu_1 \epsilon_d - \alpha)^2 > 4 (\epsilon_e^2 - \alpha \kappa )$, $\lambda_{2,3}$ are real and the stability condition becomes 
\begin{equation}
\kappa > (\alpha) \; \; \textrm{and} \; \; \epsilon_e^{2} > (\alpha \kappa  ) .
\label{eq:stability3_simpli}
\end{equation}
Thus, if Eqs.~(\ref{eq:stability1}) and (\ref{eq:stability2_simpli}) or~(\ref{eq:stability3_simpli}) are simultaneously satisfied, the oscillations can not occur and the systems stabilize to a steady state  of amplitude death.

For a given $\kappa$, $\alpha$, and $\mu_2$, the transition to amplitude death occurs at critical coupling strengths $\epsilon_{dc}$ and $\epsilon_{ec}$ independent of each other. That is 
\begin{equation}
\epsilon_{dc} = const
\label{eq:curve1}
\end{equation}
and 
\begin{equation}
\epsilon_{ec} = const
\label{eq:curve2}
\end{equation}
For different network configurations $\mu_2$ is different and the transition occurs at the critical coupling strength
\begin{equation}
\epsilon_{dc}  = \frac{-\alpha}{\mu_2} .
\label{eq:curve3}
\end{equation}

In the case where $G$ is asymmetric, using the approximation $f' \sim \alpha$  as explained above, we can write the Jacobian corresponding to Eqs.~(\ref{eq:Phi1dot}), (\ref{asym_Phikdotz}), and (\ref{eq:zdot}) as
\[ J =  \left( \begin{array}{ccc}
\alpha + \mu_2 \beta \epsilon_d & \epsilon_e z N \gamma q_{\parallel} & \epsilon_e N \gamma q_{\parallel} \\
0 & \alpha + \mu_2 \epsilon_d & \epsilon_e N \\
0 & -\epsilon_e / N  & -\kappa  \end{array} \right),\]
and the eigenvalues are the same as given in Eq.~(\ref{eq:lyap23}). Thus, we get the same stability relations as in Eqs.~(\ref{eq:stability1}), (\ref{eq:stability2_simpli}), and (\ref{eq:stability3_simpli}).

\section{Numerical analysis}
In this section, we apply our scheme to different network topologies.
First, we apply the scheme of coupling introduced in Eq.~(\ref{eq:model}) to the case of regular networks of coupled chaotic R\"ossler systems. Here, we take the coupling to be of a diffusive type [Eq.~(\ref{eq:rossler})]. The occurrence of amplitude death in the case of a regular all-to-all coupled network is illustrated in Fig.~\ref{ts:rossler}. 

To characterize the state of amplitude death, we use an index $A$ introduced in the earlier paper\cite{Res11}. It is defined as the difference between the global maximum and global minimum values of the time series of the system over a sufficiently long interval. The case where $A=0$ represents the state of amplitude death, while $A \neq 0$ indicates oscillatory dynamics.  The parameter value at which $A$ becomes $\sim 0$ is thus identified as the threshold for onset of stability of amplitude death states. 
 
\begin{figure}
\includegraphics[width=0.95\columnwidth]{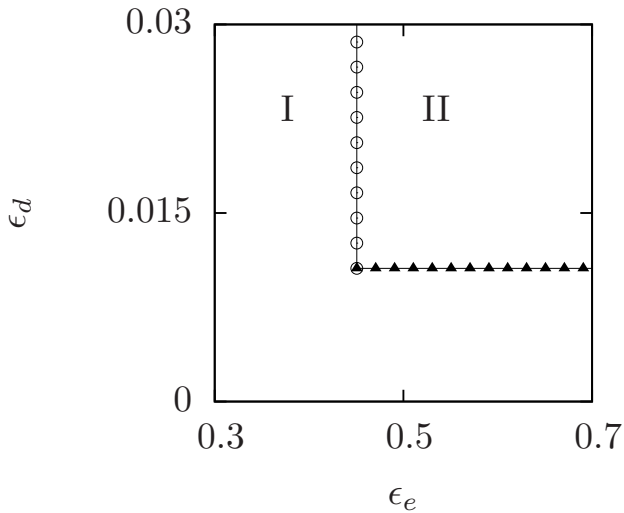} 
\caption{ \label{e1e3ros} Transition from the region of oscillations (I) to the region of amplitude death (II) is shown in the parameter plane $\epsilon_e$--$\epsilon_d$ for the coupled R\"ossler systems. Numerical simulations are done on a symmetric, all-to-all coupled network of $10$ nodes. The points mark the parameter values ($\epsilon_{ec}$,$\epsilon_{dc}$) at which the transition to the amplitude death occurs. Solid triangles show the transition to amplitude death as $\epsilon_d$ is increased for a constant $\epsilon_e$. The horizontal line formed by these triangles confirms the stability condition Eq.~(\ref{eq:curve1}). Similarly, circles correspond to transition to the amplitude death state as $\epsilon_e$ is increased for a constant $\epsilon_d$ and confirm the stability condition of Eq.~(\ref{eq:curve2}). }
\end{figure}

For a given network topology, the threshold value of coupling strengths for the onset of amplitude death is given by Eqs.~(\ref{eq:curve1}) and (\ref{eq:curve2}). This is verified for the case of an all-to-all coupled network of R\"ossler systems by direct numerical simulations. Using the index $A$ the region of amplitude death states are identified in the parameter plane of coupling strengths, $\epsilon_e$--$\epsilon_d$, and is shown in Fig.~\ref{e1e3ros}. The transition curves from the stability analysis given in Eqs.~(\ref{eq:curve1}) and (\ref{eq:curve2}) are also plotted. We see that the agreement is good.

\begin{figure}
\includegraphics[width=0.95\columnwidth]{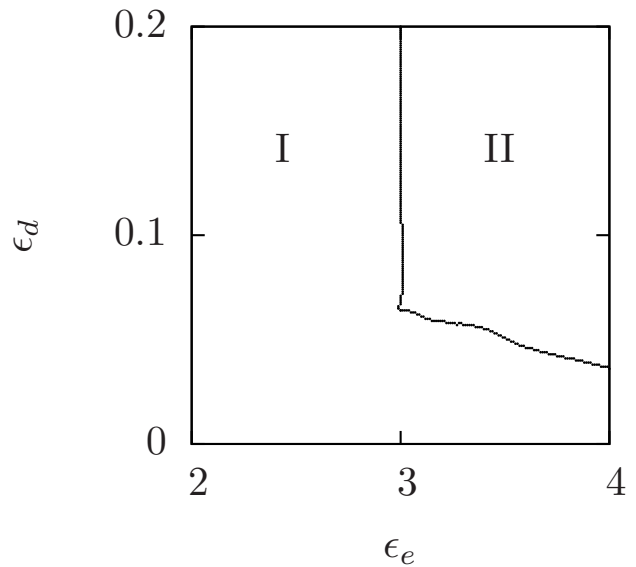}
\caption{ \label{lane1e3}  Transition from the region of oscillations (I) to the region of amplitude death (II) is shown in the parameter plane $\epsilon_e$--$\epsilon_d$ for $10$ coupled Landau systems on a symmetric, all-to-all coupled network [Eq.~(\ref{eq:landau})]. The points mark the parameter values ($\epsilon_{ec}$,$\epsilon_{dc}$) at which the transition to the amplitude death occurs. Here, the intrinsic parameter of the systems and the damping parameter of the environment are chosen to be $\omega=2$ and $\kappa=1$ respectively. We use the same set of initial conditions for any pair of ($\epsilon_e$,$\epsilon_d$) values.}
\end{figure}

However, we note that exact agreement with stability theory as shown in Fig.~\ref{e1e3ros} is seen only for R\"ossler type nodal dynamics. For a network of Landau-Stuart oscillators, given by the following equations,
\begin{eqnarray}
\dot{x}_{i1} & = & (1-x_{i1}^2 - x_{i2}^2)x_{i1} - \omega x_{i2} + \epsilon_d \sum_{j} G_{ij} x_{j1} + \epsilon_e w, 
\nonumber \\
\dot{x}_{i2} & = & (1-x_{i1}^2 - x_{i2}^2)x_{i2} + \omega x_{i1},
\nonumber \\
\dot{w} & = & -\kappa w - \frac{\varepsilon_e}{N} \sum_{i}{ x_{i1}},
 \label{eq:landau}
\end{eqnarray}
similar analysis shows some deviations between theoretical and numerical transition curves (Fig.~\ref{lane1e3}). The reason for this is the following. Towards the end of the stability analysis, we have used an approximation of constant Jacobian $f'$, which masks the system-specific details of the transition, but gives the overall features of the phase diagram. Hence the conditions (24) and (25) are approximate, and one must investigate in a specific case to see any departures from them.

For both R\"ossler and Landau-Stuart in the amplitude death state, the Jacobian $f'$ depends on $\epsilon_e$, but not on $\epsilon_d$. 
Hence, the condition (24), i.e., $\epsilon_{ec}=const$, obtained from Eq. (23) which is derived from Eqs. (13) and (14), is independent of $\epsilon_d$ for both  R\"ossler and Landau-Stuart as can be seen from Figs. 2 and 3.

The other condition (25), i.e., $\epsilon_{dc}=const$, is obtained from Eq. (21) which is derived from Eq. (15). Since Eq. (15) depends both on $\epsilon_d$ and indirectly on $\epsilon_e$ through the Jacobian $f'$, $\epsilon_{dc}$ will now depend on $\epsilon_e$ for both R\"ossler and Landau-Stuart. In the case of R\"ossler networks, the Jacobian has a simple structure and the dependence  of $\epsilon_{dc}$ on $\epsilon_e$ is weak giving almost a straight line as in Fig. 2. In the case of Landau-Stuart networks, the dependence of $\epsilon_{dc}$ on $\epsilon_e$ is a polynomial relation explaining the curve obtained for numerical simulations (Fig.~\ref{lane1e3}). 

In the context of Landau-Stuart oscillators, there is an additional complexity due to bistability with oscillations and amplitude death co-existing with different basins. This bistability has been reported earlier in the case of amplitude death in Landau-Stuart oscillators \cite{Kar07,Res11}. Such a bistability does not exist for R\"ossler systems. In Fig.~\ref{lane1e3}, we have used the same set of initial conditions for any pair of ($\epsilon_e$,$\epsilon_d$) values. For a different set of initial conditions, the critical coupling curve in Fig.~\ref{lane1e3} can shift slightly though the general features will remain the same. 
 
\begin{figure}
\includegraphics[width=0.95\columnwidth]{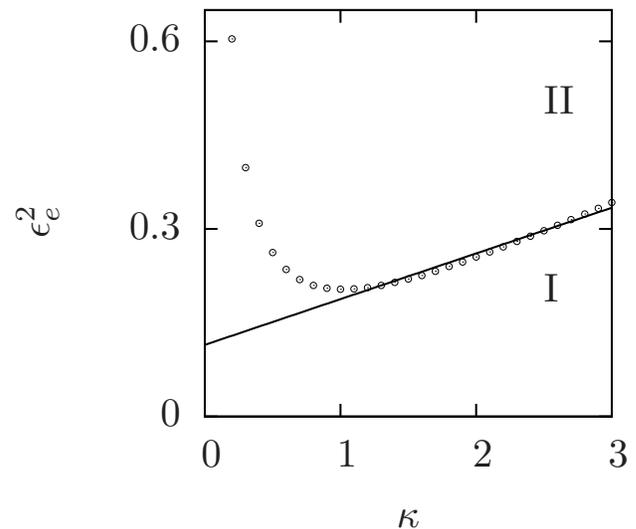}
\caption{ \label{kape1sqros} Transition from the region of oscillations (I) to the region of amplitude death (II) is shown in the parameter plane $\kappa-\epsilon_e^{2}$ for the coupled R\"ossler systems given in Eq.~(\ref{eq:rossler}). Here, an all-to-all coupled network of $10$ nodes is used. The points mark the square of the coupling strength at which the transition to the amplitude death state
occurs for each value of $\kappa$. Solid curve is a linear fit corresponding to the stability condition Eq.~(\ref{eq:stability3_simpli}). The deviation from straight line behaviour for small values of $\kappa$ is discussed in the text. The solid curve is a linear fit corresponding to the stability condition of Eq. (23). The deviation from straight line behavior for small values of $\kappa$ is discussed in the text. }
\end{figure}

We also verify numerically the criteria for transition to amplitude death given in Eq.~(\ref{eq:stability3_simpli}). n Fig.~\ref{kape1sqros}, the line corresponds to the stability conditions of Eq.~(\ref{eq:stability3_simpli}) and the points are square of the critical coupling strength $\epsilon_e$ , as $\kappa$ is varied. As we can see from Fig.~\ref{kape1sqros}, the  agreement is good for larger values of $\kappa$. However, for small values of $\kappa$, the points deviate from straight line behavior. The reason is clear from Eq.~(\ref{eq:stability2_simpli}) which gives the lower limit on $\kappa$. 

\begin{figure}
\includegraphics[width=0.95\columnwidth]{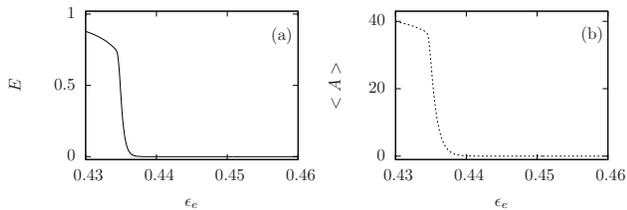}
\caption{ \label{rostransition} (a) The oscillatory part of incoherent energy, $E$, as a function of $\epsilon_e$ for a fixed value of $\epsilon_d =0.4$ for an all-to-all coupled network of $10$ R\"ossler systems. (b) The index $<A>$ as a function of $\epsilon_e$ for a fixed value of $\epsilon_d = 0.4$ for an all-to-all coupled network of $10$ R\"ossler systems. As $\epsilon_e$ is increased, we observe a continuous transition to the state of amplitude death.}
\end{figure}

 The nature of the transitions to the state of amplitude death is further characterized by fixing one of the parameters $\epsilon_e$ or $\epsilon_d$ and increasing the other. To characterize the transition, we use the oscillatory part of incoherent energy, $E$ defined in Ref.~\cite{Rub00}. Since the fixed point obtained in this case is not the origin, the oscillatory part of incoherent energy is defined after shifting the origin to the fixed point as
\begin{equation}
\label{eq:incoherent}
E = \frac{<\sum_{j=1}^{N} ( (x_{j1} - x_1^*)^2 + (x_{j2} - x_2^*)^2 + (x_{j3} - x_3^*)^2 )>}{< \sum_{j=1}^{N} ( (x_{j1}^0 - x_1^*)^2 + (x_{j2}^0 - x_2^*)^2 + (x_{j3}^0 - x_3^*)^2) >} , 
\end{equation}
where, ($x_{j1}^0$, $x_{j2}^0$, $x_{j3}^0$) represent the variables ($x_{j1}$, $x_{j2}$, $x_{j3}$) in the uncoupled case ($\epsilon_e = \epsilon_d = 0$), ($x_{1}^*$, $x_{2}^*$, $x_{3}^*$) represent the fixed point of the coupled system (Eq.~\ref{eq:rosfp}), and $<.>$ denotes average over time. In Fig.~\ref{rostransition}(a), we plot $E$ of the coupled system given in Eq.~\ref{eq:rossler} for increasing $\epsilon_e$ for a chosen value of $\epsilon_d$. Here, the transition from oscillatory state to amplitude death state is continuous such that, as the coupling strength is increased, $E$ gradually decreases to zero. However, the calculation of $E$, using Eq.~\ref{eq:incoherent} is useful only when the fixed points of the coupled system can be calculated analytically. Alternatively, we can use the index $<A>$, used to identify states of amplitude death in Figs.~\ref{e1e3ros},\ref{lane1e3} and \ref{kape1sqros}, which does not require the knowledge of the fixed point. This is shown in Fig.~\ref{rostransition}(b). Numerically, we also observe that, at each node the sub-systems undergo a reverse period-doubling bifurcation to limit cycle before undergoing a transition to the amplitude death state, similar to the case of two coupled R\"ossler systems reported earlier \cite{Res11}. Similar transition is observed for the case where $\epsilon_e$ is kept fixed and $\epsilon_d$ is increased.

\begin{figure}
\includegraphics[width=0.95\columnwidth]{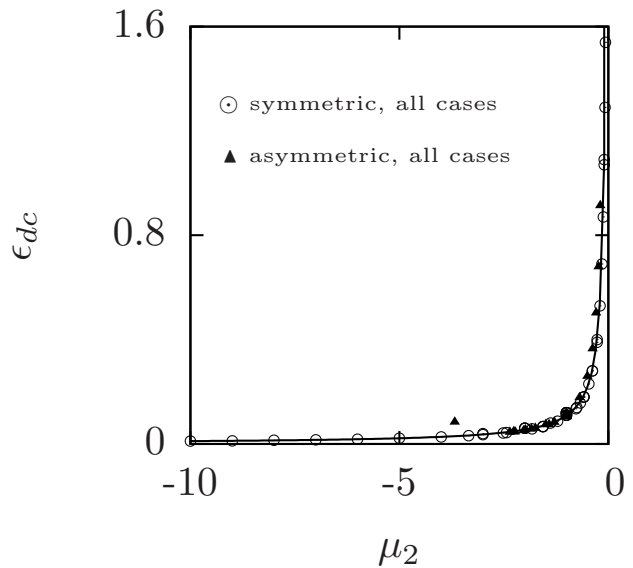}
\caption{ \label{eigep3c} Critical strength of direct coupling $\epsilon_{dc}$ for amplitude death as a function of the largest non-zero eigenvalue of the coupling matrix, $\mu_2$. Here points correspond to values obtained from numerical simulations, while the line correspond to the stability condition in Eq.~(\ref{eq:curve3}). Open circles represent symmetric networks of different topologies such as chain, ring, all-to-all-coupled, tree, lattice, star and random. Similarly, filled triangles represent asymmetric networks of different topologies such as chain, ring, tree, star and random. For the asymmetric networks, $\mu_2$ is, in general, complex. Hence, the real part of $\mu_2$ is plotted here. The parameters of the R\"ossler system are the same as that used in Fig.~\ref{ts:rossler}. The parameters in the coupling terms are $\epsilon_e = 0.8$ and $\kappa =1$.}
\end{figure}

\begin{figure}
\includegraphics[width=0.95\columnwidth]{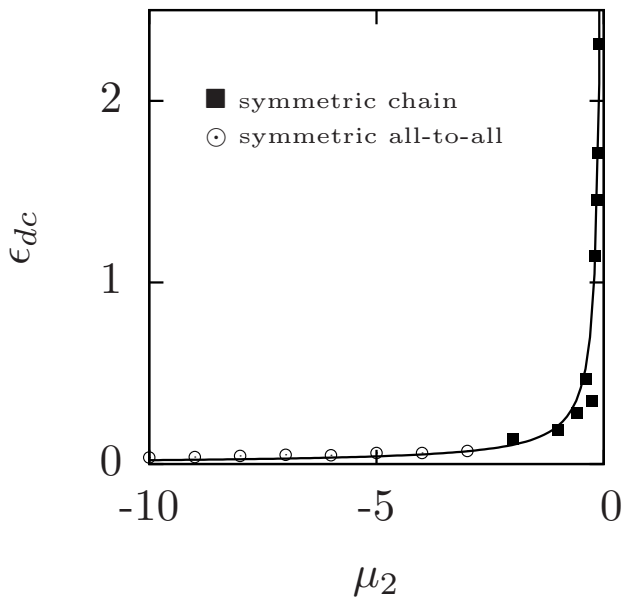}
\caption{ \label{lan_eigep3c} Critical strength of direct coupling $\epsilon_{dc}$ for amplitude death as a function of the largest non-zero eigenvalue of the coupling matrix, $\mu_2$ for Landau-Stuart oscillators. Here points correspond to values obtained from numerical simulations, while the line correspond to the stability condition in Eq.~(\ref{eq:curve3}). Open circles represent symmetric  all-to-all coupled networks of different sizes and filled squares represent symmetric networks of chain topology of different sizes. The parameters in the coupling terms are $\epsilon_e = 4.0$ and $\kappa =1$.}
\end{figure}

So far, we have presented the results from numerical simulations of Eq.~(\ref{eq:rossler}) in an all-to-all coupled network. Similar results are observed for other network configurations such as chain, ring, tree, lattice, star and random topologies. From numerical simulations of R\"ossler systems coupled in different network topologies, we see that, the critical strength of coupling via environment $\epsilon_{ec}$ is independent of the network topology. On the other hand, the critical strength of direct coupling for amplitude death, $\epsilon_{dc}$ varies with the largest non-zero eigenvalue of the coupling matrix $G$, as given in Eq.~(\ref{eq:curve3}). To verify this, we consider symmetric and asymmetric matrices of different topologies and sizes. With each network considered, the largest non-zero eigenvalue, $\mu_2$  of the corresponding coupling matrix $G$ is calculated. The critical value of coupling, $\epsilon_{dc}$ is obtained from numerical simulations of Eq.~(\ref{eq:rossler}) and is plotted against the corresponding $\mu_2$ in Fig.~\ref{eigep3c}. A universal relation between the critical coupling strength and largest non-zero eigenvalue of the coupling matrix, as given by Eq.~({\ref{eq:curve3}}) is clearly seen.
Similar behaviour is seen in the case of coupled Landau-Stuart systems on a network, as shown in Fig.~\ref{lan_eigep3c}.
A similar insensitivity of the transition to amplitude death, to the network structure is reported in the case of time-delay coupled R\"ossler systems in Ref.~\cite{Zou11} where, the smallest eigenvalue of the adjacency matrix of the network is found to determine the size of the death island.
\section{Discussion}
We report the amplitude death in complex network of nonlinear oscillators caused by interactions with a common environment. Our method involves a damped environment modeled by a single variable coupled to all nodes equally. We develop a stability analysis to obtain the criteria for the onset of amplitude death. The transition curves obtained from the stability analysis matches well with those obtained from direct numerical simulations. Moreover, the method introduced here is found to work for different network topologies. In the context of the two specific nodal dynamics, R\"ossler and Landau-Stuart studied here, we find that, there exists a universal relation which is independent of network topology, between the largest eigenvalue of the coupling matrix and the critical value of coupling. All the points corresponding to transition to amplitude death state for different network topologies fall on the same curve. This is as expected from the stability analysis developed. 

The dynamical mechanism that induces amplitude death itself is very interesting, where the environment modulates the dynamics in a self-organized way. Since amplitude death is brought about by a common variable coupled equally to all nodes, the design procedure is simple and easy to implement in cases where targeting of complex systems to steady state behaviour is desirable. 
\begin{acknowledgments}
GR was supported in part by the DST Centre for Mathematical Biology, UGC Centre for Advanced Studies and DST IRHPA Centre for Neuroscience. 
\end{acknowledgments}

\end{document}